\documentstyle[aps,multicol,prb,epsf]{revtex}
\def\d{{\rm d}}

\def\dl{{{\rm d} \over {{\rm d} \ell}} \,}
\def\dx{{\rm d}x}
\def\e{{\rm e}}
\def\virg{\;\;,}

\def\vf{v_{\rm F}}
\def\kf{k_{\rm F}}
\def\ef{\epsilon_{\rm F}}

\def\al{\alpha }

\def\gnb{\tilde{g}_1}
\def\gnf{\tilde{g}_2}
\def\gtb{\tilde{g}_1}
\def\gtf{\tilde{g}_2}
\def\tt{\tilde{t}}
\def\gt{\tilde{g}}
\def\gtu{\tilde{g}_3}
\def\gtuc{\tilde{g}_{3 {\rm c}}}

\def\lg{l_{\rm g}}

\def\om{\omega_{\rm m}}
\def\ggs{\buildrel\textstyle > \over {\hbox{\raise0.2ex\hbox{$\sim$}}}}
\def\lls{\buildrel\textstyle < \over {\hbox{\raise0.2ex\hbox{$\sim$}}}}
\def\gsim{\,\lower0.75ex\hbox{$\ggs$}\,}
\def\lsim{\,\lower0.75ex\hbox{$\lls$}\,}
\def\Krho  {K_{\rho}}
\def\Ksigma{K_{\sigma}}
\def\Kc   {K_{\rm C}}
\def\Ks   {K_{\rm S}}
\def\gncp {G_{\nu + ,{\rm C} +}}
\def\gncm {G_{\nu + ,{\rm C} -}}
\def\gnc  {G_{\nu + ,{\rm C} p}}
\def\gnsp {G_{\nu + ,{\rm S} +}}
\def\gnsm {G_{\nu + ,{\rm S} -}}
\def\gns  {G_{\nu + ,{\rm S} p}}
\def\gcs  {G_{{\rm C} p ,{\rm S} p'}}
\def\gcsp {G_{{\rm C} p ,{\rm S} +}}
\def\gcsm {G_{{\rm C} p ,{\rm S} -}}
\def\gcps {G_{{\rm C} + ,{\rm S} p}}
\def\gcms {G_{{\rm C} - ,{\rm S} p}}
\def\gcpsp{G_{{\rm C} + ,{\rm S} +}}
\def\gcpsm{G_{{\rm C} + ,{\rm S} -}}
\def\gcmsp{G_{{\rm C} - ,{\rm S} +}}
\def\gcmsm{G_{{\rm C} - ,{\rm S} -}}
\def\grc  {G_{\rho +, {\rm C} p}}
\def\grcp {G_{\rho +, {\rm C} +}}
\def\grcm {G_{\rho +, {\rm C} -}}
\def\grs  {G_{\rho +, {\rm S} p}}
\def\grsd {G_{\rho +, {\rm S} p'}}
\def\grsp {G_{\rho +, {\rm S} +}}
\def\grsm {G_{\rho +, {\rm S} -}}
\def\gpc  {G_{\sigma +, {\rm C} p}}
\def\gpcp {G_{\sigma +, {\rm C} +}}
\def\gpcm {G_{\sigma +, {\rm C} -}}
\def\gps  {G_{\sigma +, {\rm S} p}}
\def\gpsd {G_{\sigma +, {\rm S} p'}}
\def\gpsp {G_{\sigma +, {\rm S} +}}
\def\gpsm {G_{\sigma +, {\rm S} -}}
\begin{document} 
\draft

\title
{
Confinement of Interchain Hopping 
  by Umklapp Scattering   in Two-Coupled  Chains 
 }

\author{
 Y.  Suzumura and  M.  Tsuchiizu 
}

\address{
Department of Physics, Nagoya University, Nagoya 464-8602, Japan \\
}
\author{
  G.  Gr\"uner
}

\address{
Department of Physics, University of California, Los Angeles, CA 90095-1547, USA
}

\date{\hspace{3.5cm} }
\maketitle

\begin{abstract}
 The effect of umklapp scattering on interchain hopping has 
 been investigated for  two-coupled chains of interacting electrons
 with half-filled band. 
 By analyzing in terms of  renormalization group method,  we have 
found that interchain hopping is renormalized  to zero 
 and is   confined  when a gap induced by 
 umklapp scattering   becomes larger than a critical  value. 
 From a phase diagram calculated on a plane of the interchain hopping 
 and the gap,  we discuss a role of the correlation gap 
 which has been studied in  metallic state at temperatures 
 above spin density wave state 
 in organic conductors. 
\end{abstract}


\begin{multicols}{2}
 The linear chain conductors, called Bechgaard salts and described 
 by the formula (TMTTF)$_2$X and (TMTSF)$_2$X - where TMTTF and TMTSF 
stands for tetramethyltetrathiofulvalene and tetramethyltetraselenofulvalene respectively, and X refers to various counter ions - 
 have been, over the years the subject to intensive studies. 
 Early attention has focused on the various broken symmetry 
 (magnetic and superconducting) states but recently the state 
 above  the phase transition became the subject of 
 intensive studies. 
 In these salts the transfer integrals are different 
 in different directions and 
they span a wide  range of dimensionality
\cite{Bechgaard}. 
 While the band width along the chain direction is comparable 
 in the various salts, 
 and the bandwidth in the least conduction direction is rather small, 
 the transfer integral in the second best conducting ($b$) direction 
 increases going from the TMTTF to the TMTSF salts
\cite{Jerome,Grant}.
 One central feature of these  salts is that there is a transfer of 
 one electron from the TMTTF or TMTSF chain to the X counter ions, 
 and also there is a dimerization along the TMTTF and TMTSF chains 
 and thus these materials can be regarded as having a half-filled 
electron band, thus umklapp scattering is important. 

 Various experiments give evidence for a charge gap for the TMTTF salts  and a metallic behavior for the TMTSF salts. 
 Recent optical, transport and dielectric experiments
\cite{Gruner_ICSM}, 
 taken together with photoemission measurement
\cite{Gruner_prepri} lead to a picture 
where, with increasing transfer integral $t_b$ a transition 
 occurs from an insulating state where electrons are confined 
 to the individual chains, 
to a metallic state where the electrons are deconfined. 
 This transition occurs where $t_b$ becomes comparable 
 to the charge gap. 

 These conductors have been studied theoretically by use of a model 
of  quasi-one-dimensional electron systems 
 having repulsive intrachain interaction without umklapp scattering.  
  The  hopping perpendicular to chain 
  becomes relevant   even for a small transfer energy
\cite{Yakovenko} 
 although the hopping is suppressed by one-dimensional fluctuation
\cite{Bourbonnais}.
 Two-coupled chains has been studied as a basic model which  includes 
  intrachain interaction  and transverse hopping. 
   For the Tomonaga-Luttinger model with   forward 
   scattering , it has been shown that  
 a  gap appears in  the  transverse  density fluctuations 
 and that   degeneracy of  in-phase and out-of-phase pairings 
 of density waves   is removed
\cite{Nersesyan,Fabrizio,Finkelstein,Yoshioka_JLTP}. 
 The model with backward scattering exhibits a phase diagram 
 which is  different from that of a single chain. 
 In case  of  Hubbard model with   a repulsive interaction 
and a incommensurate band,    
 the ground state of two chains is given by   the  SC state with 
 interchain and in-phase pairing, i.e.,  
 $d$-wave like paring\cite{Fabrizio,Schulz,Balents}. 
  The transverse hopping   becomes 
   relevant  even for a small transfer energy 
 unless the intrachain interaction is extremely large\cite{Fabrizio}.
 On the other hand,  it has been maintained that 
 confinement  with no coherent single particle hopping 
occurs  in  coupled chains  of Luttinger liquids 
 for the interchain hopping smaller than a critical value
\cite{Anderson,Clarke,Strong}.

 In this letter,  two-coupled chains 
 with intrachain interaction and  half-filled band 
 is considered. 
 The model applies to  the normal state of 
  organic conductors,  TMTSF and TMTTF salts,
 for which    the importance of umklapp scattering 
 has been pointed out also earlier\cite{Emery,Giamarchi_physica}.   
  We   demonstrate a novel fact  that 
 the interchain hopping becomes irrelevant and confined 
 with increasing  the magnitude of umklapp scattering.  
 The relevance of our result to   experiments 
  is also discussed.  


We consider  two-coupled chains  with the intrachain interaction 
 and  interchain electron hopping.
 The kinetic energy  parallel to the chain 
 is linearized with the Fermi velocity 
 $\vf$ (-$\vf$) and  Fermi momentum $\kf$ for 
 the right-going  (left-going) electron, respectively.  
 The intrachain interactions consist of   
 forward scattering, backward scattering and umklapp scattering 
 whose coupling constants are defined as 
$g_2$,  $g_1$  and $g_3$ respectively. 
 After diagonalization of  the term for interchain hopping,  
 the kinetic energy is expressed  in terms of  bonding state 
  and antibonding state with  new  Fermi momentum, 
    $k_{{\rm F}\pm} \equiv \kf + (\mp t/\vf)$, 
 where $t$ denotes a  hopping energy.  
 Applying  the  bosonization  method to electrons  around the  
new Fermi points, 
we introduce  Bose fields of  phase variables,  
  $\theta_{\rho +}$ and  $\theta_{\sigma +}$
($\theta_{{\rm C}+}$ and $\theta_{{\rm S}+}$),  
 which  express   fluctuations for 
 the total (transverse) charge density and 
  spin density  respectively.\cite{Yoshioka_JLTP}
 The commutation relation with  conjugate phase  
is   given by  
$ 
  [\theta_{\rho   +}(x),\theta_{\rho   -}(x')]
 =[\theta_{\sigma +}(x),\theta_{\sigma -}(x')] 
 =[\theta_{{\rm C}+}(x),\theta_{{\rm C}-}(x')]
 =[\theta_{{\rm S}+}(x),\theta_{{\rm S}-}(x')]
 = i \pi {\rm sgn}(x-x')
 $.
 In terms of  these phase variables and  
 the  bosonization for the field operator\cite{Luther}, 
 our   Hamiltonian is given by
\end{multicols}
\begin{eqnarray} \label{eqn:H}
{\cal H} &=& 
\sum_{\nu = \rho,\sigma} \frac{v_\nu}{4\pi} \int \dx
 \left\{
   \frac{1}{K_\nu} \left(\partial \theta_{\nu_+} \right)^2
         +  K_\nu  \left(\partial \theta_{\nu_-} \right)^2
 \right\}
+
\sum_{\nu = {\rm C},{\rm S}} \frac{\vf}{4\pi} \int \dx
 \left\{
   \frac{1}{K_\nu} \left(\partial \theta_{\nu_+} \right)^2
         +  K_\nu  \left(\partial \theta_{\nu_-} \right)^2
 \right\}          
\nonumber \\
&+& \hspace{-1mm} 
    \frac{2g_2 - g_{1}}{4\pi^2 \al^2} \hspace{-1mm} \int \dx 
    \left\{ 
        \cos \left( \sqrt{2}\theta_{{\rm C}+} - 4tx/\vf \right) 
      + \cos \sqrt{2} \theta_{{\rm C}-}  
    \right\} 
    \left\{
        \cos \sqrt{2} \theta_{{\rm S} +} - \cos \sqrt{2} \theta_{{\rm S} -}
    \right\} \nonumber \\
&+& \hspace{0.5mm} 
    \frac{-g_{1}}{4\pi^2 \al^2}\hspace{1mm}  \int \dx
    \left\{ 
        \cos \left( \sqrt{2}\theta_{{\rm C} +} - 4tx/\vf  \right) 
      - \cos \sqrt{2} \theta_{{\rm C} -}  
    \right\} 
    \left\{
        \cos \sqrt{2} \theta_{{\rm S} +} 
             + \cos \sqrt{2} \theta_{{\rm S} -}
    \right\} \nonumber \\
&+& 
    \frac{g_{1}}{2\pi^2 \al^2}  \int \dx
    \cos \sqrt{2} \theta_{\sigma +} 
    \left\{
        \cos \left( \sqrt{2}\theta_{{\rm C} +} - 4tx/\vf \right)
      - \cos \sqrt{2} \theta_{{\rm C} -} 
      - \cos \sqrt{2} \theta_{{\rm S} +} 
      - \cos \sqrt{2} \theta_{{\rm S} -} 
    \right\}            \nonumber \\
&+& 
    \frac{g_{3}}{2\pi^2 \al^2} \int \dx
    \cos \sqrt{2} \theta_{\rho +} 
    \left\{
        \cos \left( \sqrt{2}\theta_{{\rm C}+} - 4tx/\vf \right)
      + \cos \sqrt{2} \theta_{{\rm C} -} 
      - \cos \sqrt{2} \theta_{{\rm S} +} 
      + \cos \sqrt{2} \theta_{{\rm S} -} 
    \right\}  ,      
\label{phase_Hamiltonian}
\end{eqnarray}
\begin{multicols}{2}
\noindent
where
$ v_\rho = \vf \sqrt{1 - (2\gnf-\gnb)^2}$, 
 $v_\sigma = \vf \sqrt{1 - \gnb^2}$,
$\Krho = [\{1  -(2\gnf - \gnb)\}
                 /\{1 + (2\gnf - \gnb)\}]^{1/2}$, 
$\Ksigma = [ (1  + \gnb)/(1  - \gnb)]^{1/2}$ and  
$\Kc = \Ks = 1$. 
 The quantity $\alpha(\sim 1/\kf)$  
 is of the order of the lattice constant 
 and $\gt_j = g_j/(2 \pi \vf) $ with $j$ =1,2 and 3. 
 In deriving Eq. (\ref{eqn:H}), 
   a phase factor of the bosonized field operator, 
which is added to retain the aniticommutation relation, 
  is taken  so as to  conserve the sign of interaction
\cite{T2}.   

We reexpress the nonlinear term 
in Eq. (\ref{phase_Hamiltonian})  as
 $ (\vf/\pi \alpha^2)
 G_{\nu p,\nu' p'} \cos \sqrt{2} \bar{\theta}_{\nu p} 
                    \cos \sqrt{2} \bar{\theta}_{\nu' p'} $ 
 where $\sqrt{2} \bar{\theta}_{\nu p} = 
        \sqrt{2}\theta_{\nu p} -4tx/\vf$ for $\nu ={\rm C}$ and $p=+$,
 and  $\sqrt{2} \bar{\theta}_{\nu p} 
     = \sqrt{2}\theta_{\nu p} $ otherwise. 
 In the present case, there are   twelve  coupling constants,
 which are  given by 
$
\gcpsp =   \gnf-\gnb 
$,
$
\gcpsm = - \gnf 
$,
$
\gcmsp =   \gnf
$,
$
\gcmsm = - \gnf + \gnb 
$,
$
\gpcp = -\gpcm = -\gpsp = -\gpsm = \gnb 
$ and
$
\grcp = \grcm = -\grsp = \grsm = \widetilde{g}_{3}
$.
 The renormalization group method is applied to 
   response functions
for  SDW, 4$\kf$-CDW and   SC states, which  are calculated with  
 the assumption that  
  response function are scaled to  the same form  for 
  $\alpha \to \alpha '=\alpha \e^{\d l}$ 
 \cite{Giamarchi_JPF,Giamarchi_PRB}. 
 Thus   renormalization group equations   
  within the second order are obtained as 
( $\nu = \rho, \sigma$ and  $p,p' = \pm$) 
\begin{eqnarray}    \label{K:nu}
\dl K_\nu &=& 
  - \frac{1}{2 \tilde{v}_\nu^2} 
   K_{\nu}^2 
  \left[
    \gncp^2 \, J_0(y) + \gncm^2 
\right. \nonumber \\ && \left. \hspace{2cm}
     + \gnsp^2 + \gnsm^2 
  \right]
,\\
          \label{K:Theta}
\dl \Kc &=& 
  \frac{1}{2} \sum_{p=\pm} 
   \left[
     \left( - \Kc^2 \, J_0(y) \, \delta_{p,+} 
                                         + \delta_{p,-} \right)
\right. \nonumber \\ && \left. \hspace*{-1cm} \times
       \left\{ \gcsp^2 + \gcsm^2 + \grc^2 + \gpc^2 \right\}
   \right]
,\\
          \label{K:Phi}
\dl \Ks &=&
  \frac{1}{2} \sum_{p=\pm} 
   \left[
       \left( - \Ks^2 \, \delta_{p,+} + \delta_{p,-} \right)
\right. \nonumber \\ && \left. \hspace*{0cm} \times
       \left\{ 
          \gcps^2 \, J_0(y) + \gcms^2 
\right.\right. \nonumber \\ && \left.\left. \hspace*{2cm}
            + \grs^2 + \gps^2 
       \right\} 
   \right]
,\\
          \label{G:Theta}
\dl \gnc &=& 
    \Bigl( 2 - K_\nu - \Kc^{p} \Bigr) \gnc
 \nonumber \\ &&  \hspace*{-.5cm}
      - \gnsp \gcsp - \gnsm \gcsm 
 \virg  \\
          \label{G:Phi}
\dl \gns &=& 
    \Bigl( 2 - K_\nu - \Ks^{p} \Bigr) \gns
 \nonumber \\ &&  \hspace*{-1cm}
      - \gncp \gcps \, J_0(y) - \gncm \gcms
  \virg  \\
          \label{G:TL}
\dl \gcs &=& 
    \left( 2 - \Kc^p - \Ks^{p'} \right) \gcs
 \nonumber \\ &&  \hspace*{-1cm}
    - \frac{1}{\tilde{v}_\rho} \, \grc \grsd 
    - \frac{1}{\tilde{v}_\sigma} \, \gpc \gpsd 
 \virg \\
          \label{G:t}
\dl \tt(l) &=& 
    \tt(l) - \frac{1}{8} 
      \left( 
         \gcpsp^2 + \gcpsm^2 
\right. \nonumber \\ && \left. \hspace{.5cm}
         + \grcp^2 + \gpcp^2
      \right)  \Kc \, J_1(y)  \virg
\end{eqnarray}
 where  
  $K_{\nu}^{p} = K_{\nu}^{\pm 1}$ for $p = \pm$,  
 $\tilde{v}_{\nu}=v_{\nu}/\vf$,
$\tt (l) = t (l)/(\vf \alpha^{-1})$, 
 $y = 4\tilde{t}(l)$ 
 and $J_n(y)$, $(n=0,1)$, is the Bessel function. 
  The variable,  $l$,  is written explicitly 
 only for  $\tilde{t}(l)$  where  
$
\tilde{t}(0) = t/\ef \equiv \tt 
$ 
 with  $\ef = \vf/\alpha$   
 and the corresponding energy is given by  $ \ef \exp [-l]$. 
 Note that these equations in the zero limit of $t$  
 becomes equal to those of one-dimensional case\cite{Solyom}. 


 We examine both cases of  
$\tilde{g}_1 = \tilde{g}_2 \not= 0 $  and 
 $\tilde{g}_1 = 0, \tilde{g}_2 \not = 0 $  
 by calculating   renormalization group equations 
 for $K_{\rho}(l)$, $K_{\sigma}(l)$, $K_{\rm C}(l)$, 
 $K_{\rm S}(l)$   and $G_{\nu p,\nu' p'}(l)$  
 with several choices of $\gtf$, $\gtb$, $\gtu$ and $\tt$.
 For the relevant interchain hopping, 
  $\tt(l)$  increases rapidly  with increasing $l$ 
 while   $\tt(l)$  decreases to zero for  the irrelevant  hopping.  
  The relevant $\tt (l)$ corresponds to  
 $K_{\rm C}(l) \rightarrow \infty $ which comes from 
 the rapid oscillation of $J_0(y)$ 
 in  Eq. (\ref{K:Theta}).   
 The quantity $K_{\rm C}(l)$ represents 
 the degree of transverse charge fluctuation.  
 Thus    deconfinement (confinement) is obtained when  
 the limiting value of $ K_{\rm C}(l)$ 
  becomes  infinite (finite). 

 In Fig. 1, $\tt (l)$ and $1/K_{\rm C}(l)$  
  are shown as a function of $l$ 
by  solid  curve and dotted curve, respectively 
 with the fixed $\gtu=$ 0.1, $\gtuc$ (=0.189) and 0.3   
 where $\tt =0.1$ and $\gt_1 = \gt_2  = 0.3$.  
  The case for  $\gtu$ =0.1 (curves (1) and (4)) 
\begin{figure}[t]
%
\epsfxsize=3.2in\epsfbox{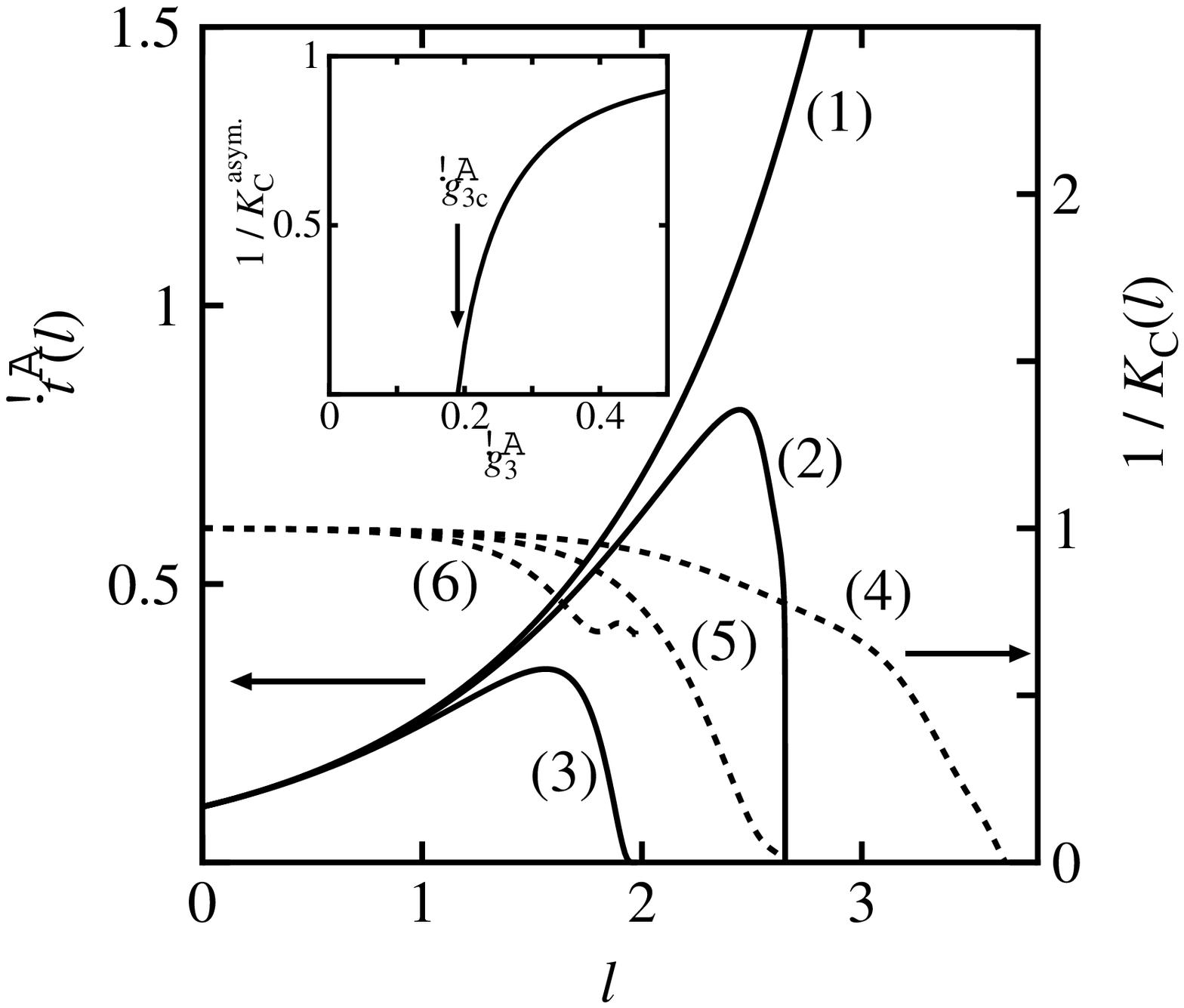}

{\noindent Fig. 1:
The $l$-dependences of  $ \tt (l)$  
 and   $1/K_{\rm C}(l)$  
 are shown by  solid curve and  dotted curve respectively 
 for   $\gtu$= 0.1 ((1) and (4)), 
   $\gtu = \gtuc$ $(=0.189)$ ((2) and (5)) and  
     $\gtu = 0.3$ ((3) and (6)), respectively   
 where   $\tt=0.1$ and   $\gtb = \gtf = 0.3$. 
  The inset shows  the $\gtu$-dependence of
  $1/K_{\rm C}^{\rm asym.}$ which corresponds to 
 the limiting value of $1/K_{\rm C}(l)$.  
}
\end{figure}
\noindent
\begin{figure}[t]
\vspace*{-.5cm}
\epsfxsize=2.9in\epsfbox{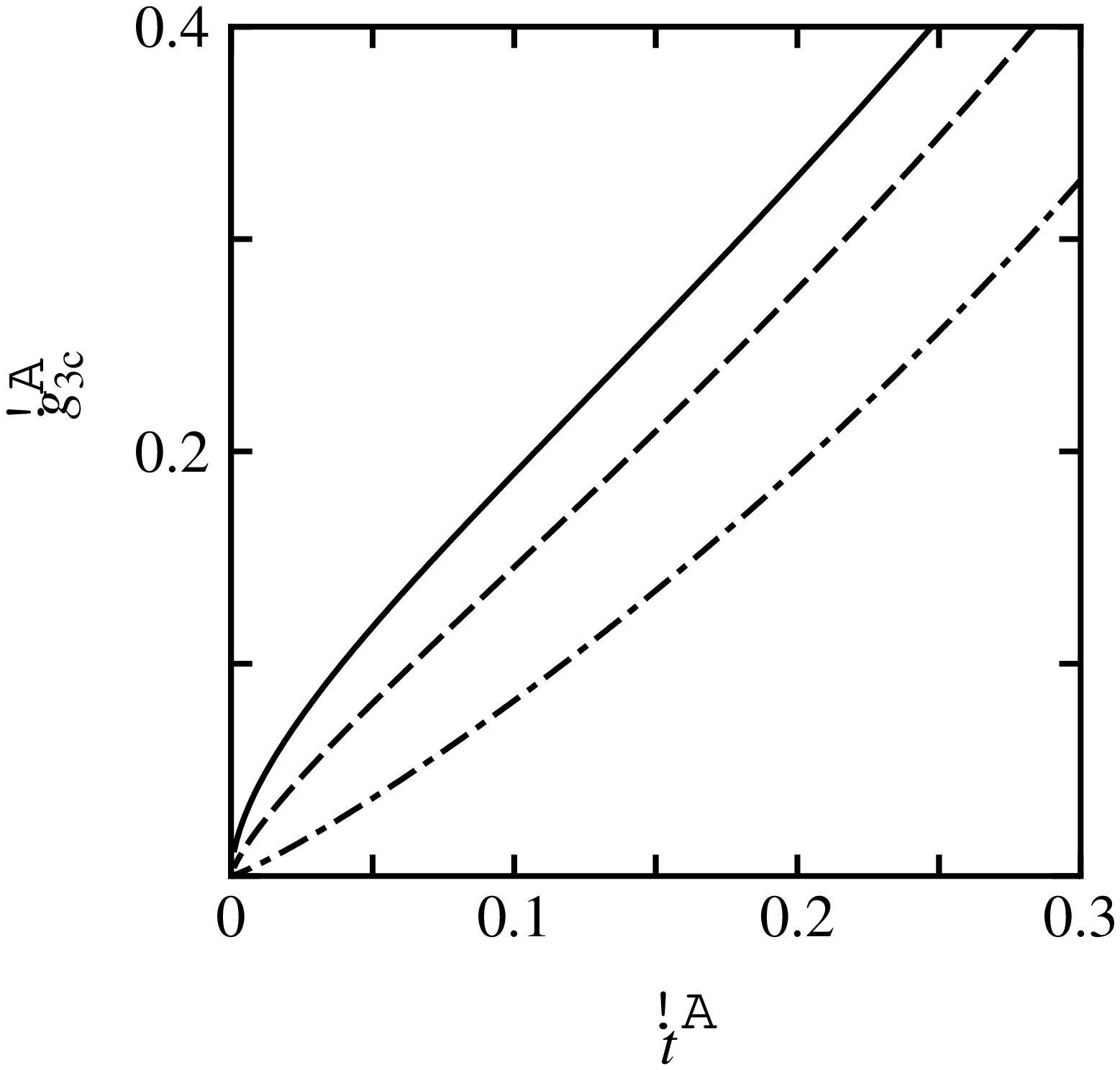}
%
{\noindent Fig. 2:
 The $\tt$-dependence of $\gtuc$ for 
 $\tilde{g_1} = \tilde{g_2} =0.3$ (solid curve), 
 $\tilde{g_1} = \tilde{g_2} =0.4$ (dashed   curve) 
 and  $\tilde{g_1} = 0$, $\tilde{g_2} =0.3$ (dash-dotted curve).   
  The case of $\gtu > \gtuc$ ($\gtu < \gtuc$) 
corresponds to confinement (deconfinement). 
}
\end{figure}
\vspace*{.1cm}
\noindent
  shows the result leading to  
  deconfinement. 
 With increasing $l$,   
   $\tt (l)$  increases rapidly 
 and $1/K_{\rm C}(l)$  decreases  monotonically to zero.  
 Our solution stops at a value of $l$ corresponding to 
 $ \Krho (l) \simeq 0$ due to the divergence of some of 
 $G_{\nu p,\nu' p'}(l)$ 
 since the present treatment is of the second order 
 for the renormalization group equations.  
 The  case for  $\gt_3 =  0.3$ (curves (3) and (6)) 
 shows a typical behavior  for  confinement.   
  With increasing $l$,    $\tt (l)$  reduces to zero 
 after taking a maximum 
 and   $1/K_{\rm C}(l)$  remains finite even 
 at the limiting value of $l$.
 There is  a crossover from  deconfinement to confinement 
  around the location of $l$ corresponding to   
 the  maximum of $\tt (l)$ 
 where $G_{\rho +, {\nu p}}(l)$ becomes of the order of unity.   
  We also obtained  that 
 $G_{\rho +,{\rm C +}}(l)/G_{\rho +,{\rm C -}}(l)
 \simeq 1/K_{\rm C}(l)$ for the limiting value,  
 indicating    the irrelevance of 
 the misfit parameter and then  the interchain hopping. 
 For  a critical value given by  
  $ \gt_3 = \gtuc$ (curves (2) and (5)),
 one finds a marginal behavior  where      
 both $\tt (l)$  and   $1/K_{\rm C}(l)$  reduce to zero 
 at the limiting  value of $l$. 
 The $l$-dependence of $K_{\rm C} (l)$ indicates that 
 there is   a transition from  deconfinement to  confinement 
 as a function of $\gtu$ 
 in the limit of low energy. 
In the inset, 
 the $\gtu$-dependence of  $1/K_{\rm C}^{\rm asym.}$ is shown 
  where  $K_{\rm C}^{\rm asym.}$ is   
  the limiting value of  $K_{\rm C}(l)$.  
 The location of $\gtuc$ is shown by the arrow.   
  For most  parameters   leading  to  
 $\gtu = \gtuc$,  the present calculation shows 
  common feature  that   
  a peak height of  $\tt (l)$  is  about  0.82 
 and   $ \om / t \simeq 0.94$ where 
 $\om$ is the energy at  the peak of $\tt(l)$.  
 We note that the Bessel function, $J_1(y)$, in r.h.s. of 
Eq. (\ref{G:t}) plays a crucial role to obtain such a transition 
 where the effect of second term of Eq. (\ref{G:t}) 
 is negligible for the relevant $\tt (l)$ and 
 becomes large for the irrelevant  $\tt (l)$.  
 With increasing $l$, $K_{\rho}(l)$ decreases  to zero 
 where  a charge gap is formed
 for $K_{\rho}(l) \simeq K_{\rho}(0)/2$, {\it e.g.},  
 at  $l \simeq 3.25 (1.50)$ for $\gtu =0.1 (0.3)$. 
  The quantity  $K_{\rm S}(l)$ 
 corresponding to  transverse spin fluctuation is also 
 suppressed by umklapp scattering. 
  The behavior of total spin fluctuation  indicates 
 the absence of the spin gap even at low energies  
 since    $K_{\sigma}(l)$  
 is almost the same as    one-dimensional one.
    Thus one finds  that 
 there is a  separation of freedoms of charge and spin   
  at energy corresponding to  a correlation gap.  
Note that the decreases of $K_{\sigma}(l)$ and $K_{\rm S}(l)$ 
 are attributable to  the backward scattering. 
In fact,   $K_{\sigma}(l)=K_{\rm S}(l)=1$ 
 for both regions of confinement and deconfinement  when $\gtb=0$.

In Fig. 2, 
 the  $\tt$-dependence of  $\gtuc$ is shown 
 for $\gtf = \gtb = 0.3$ (solid curve),  
  $\gtf = \gtb = 0.4$ (dashed  curve) and   
  $\gtf = 0.3, \gtb = 0$ (dash-dotted curve) where    
 the region for confinement (deconfinement) 
 is given by  $\gtu > \gtuc$ ($\gtu < \gtuc$).  
 The boundary is determined mainly by the competition between 
  umklapp scattering and   interchain hopping.  
 In addition to $\gtu$, 
  both $\gtf$ and $\gtb$ enhance the region for  confinement 
  where the effect of the 
forward scattering is larger than the backward scattering.  
As $\tt$ goes to zero,   $\gtuc$ reduces to zero and then 
  the confinement does not exist  in the absence of  
  umklapp scattering.

 Now we examine   the correlation  gap, $\Delta$,  
  defined by  
$
 \Delta \equiv \ef \exp [- \lg]
$ where  
  $\lg$ is evaluated from   
$
 K_{\rho}(\lg)=K_{\rho}(0)/2$.  
 We note  that  such a definition of gap well reproduces 
 a magnitude of  gap for the one-dimensional Hubbard model 
 with the weak coupling\cite{Woynarovich}. 
 It is found that  
 $\Delta$ is slightly larger than  the energy, $\om$,    
  corresponding  to  a peak  of $\tt(l)$ in Fig. 1. 
 In the inset of Fig. 3, $\Delta$ is shown as a function of 
 $\gtu$ 
for $\gtf=\gtb=0.3$ (1),  $\gtf=\gtb=0.4$ (2) and  
$\gtf=0.3, \gtb=0$ (3) with the fixed $\tt=0.1$. 
  The quantity $\Delta$, which is determined 
 mainly by $\gtu$, is enhanced also by $\gtf$ and $\gtb$.    
 The $\tt$-dependence of  $\Delta$ is  small 
 as is seen from  curve (4) which is calculated
\begin{figure}[t]
%
\epsfxsize=2.9in\epsfbox{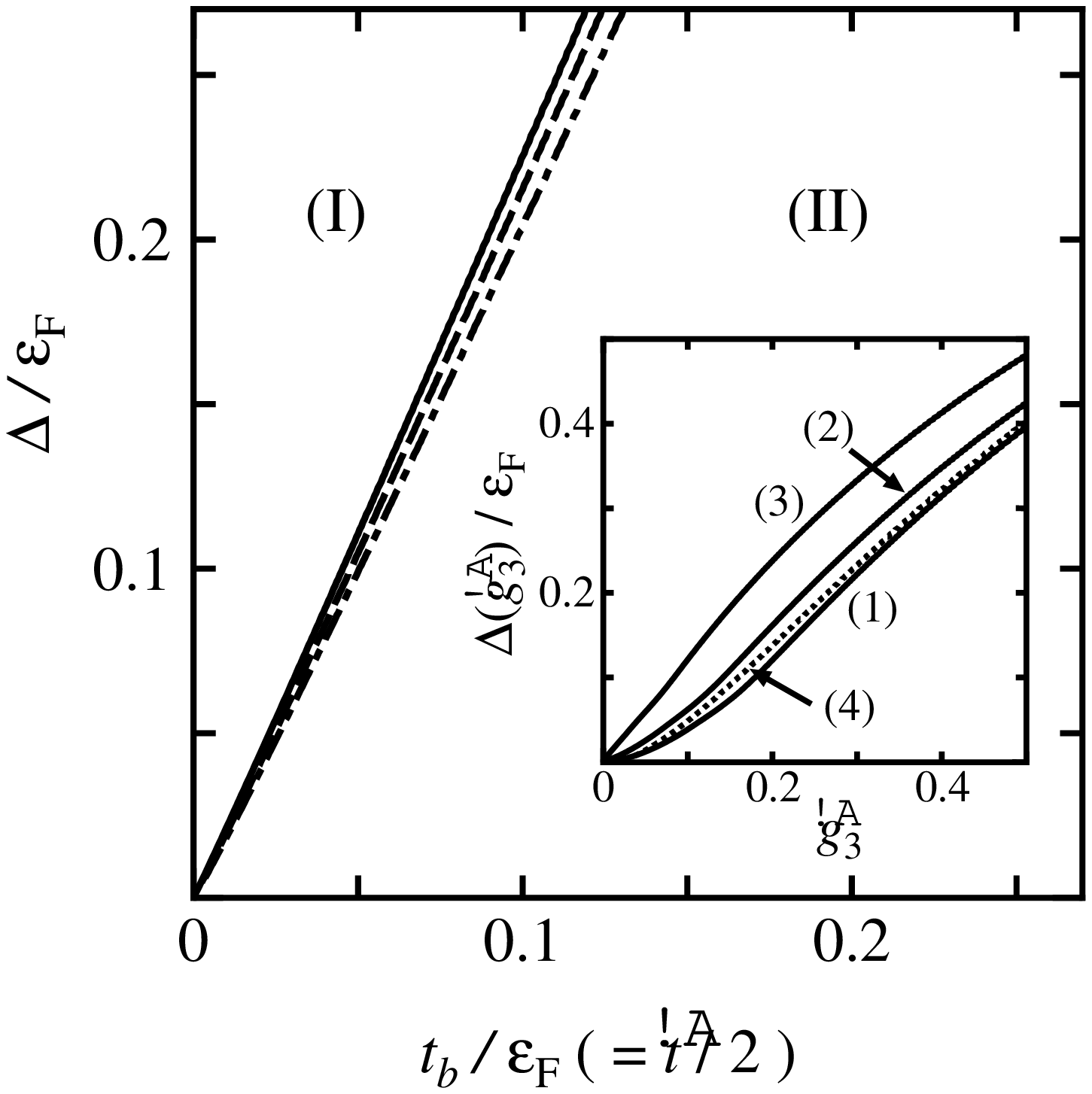}
%
{\noindent Fig. 3:
  The phase diagram of  confinement (region (I)) 
   and  deconfinement (region (II)) on the plane of 
  the interchain transfer energy, $t_b (=\tt/2)$ and 
  the correlation gap, $\Delta$. 
 The solid, dashed and dash-dotted curves denote boundaries 
 which are obtained from respective curves in Fig. 2.  
 In the inset, the correlation gap, $\Delta$ 
 is shown as a function of $\gtu$ 
 for $\gtf=\gtb=0.3$, $\tt=0.1$ (1),
  $\gtf=\gtb=0.4$, $\tt=0.1$ (2), 
  $\gtf=0.3$, $\gtb=0$, $\tt=0.1$ (3) and 
  $\gtf=\gtb=0.3$, $\tt=0.01$ (4), respectively. 
}
\end{figure}
\vspace*{.1cm}
\noindent
for $\gtf=\gtb=0.3$ and  $\tt=0.01$. 
 Here we  introduce $t_b$ defined as  
  the transfer energy perpendicular 
 to chain  for quasi-one-dimensional system
  where  $t_b = t/2$ from the definition of our Hamiltonian. 
 In terms of  
 $ \Delta$ and $t_b (= t/2)$, 
 the phase diagram is shown in Fig. 3  
 where  region (I) and region (II) correspond to 
 confinement and deconfinement, respectively. 
 Three  boundaries given by the solid curve, 
 the  dashed curve and the dash-dotted curve are evaluated  
 from   the corresponding curves in Fig. 2.  
 It turns out that the ratio of the correlation gap 
 to the perpendicular transfer energy is 
  $\Delta/t_b = 1.8 \sim 2.3$  for the interval range of 
 $ 0.01 < t_b / \ef < 0.1$.  
 This value is in excellent agreement with experiments
\cite{Gruner_prepri}  which indicate a transition from 
 a confined insulator to a deconfined metal for between 1.5 and 2. 
 The critical value  of $\Delta$ for the  confinement 
 decreases  for the large  $\gtf$ and $\gtb$.


 The  dominant state, which is found  with decreasing 
$\omega (= \ef \exp [-l])$ and for the fixed   $\gtu$  and 
 $\gtf = \gtb >  0$,
 is examined by calculating  response functions for  
 SDW   with  the intrachain and  out-of-phase   pairing and    
  4$\kf$-CDW  with  the intrachain and  in-phase   pairing   
 and SC state with the interchain and in-phase pairing. 
   When $\Delta \gsim  t $ (i.e., $\gtu > \gtuc$), 
 there is a  crossover from deconfinement to confinement  
  in SDW state at energy given by 
  $\omega \simeq \om (< \Delta)$. 
 Further, the SDW state moves into the  confined  4$\kf$-CDW state 
 at lower energies. 
   When $\Delta \lsim  t$,  
 all the  states are deconfined  and 
 SDW state  is replaced by  4$\kf$-CDW state  
 at energy  much lower than $\Delta$. 
 The  SC state is possible  for the region 
  of deconfinement with  $\gtu \ll \tt$ and finite energy.   
 We note that  SC state is also  found in  other region of  
  $2 \gtf-  \gtb < - |\gtu|$, 
where the umklapp scattering becomes  irrelevant\cite{Emery}.  
  
 In conclusion, we have found, 
 by  examining  the effect of umklapp scattering
  on  the interchain hopping in  two-coupled chains, 
 that the interchain hopping becomes irrelevant 
 resulting in 
 the  transition from deconfinement to   confinement  
 when correlation gap induced by umklapp scattering  
becomes larger than   the interchain hopping. 
 This result supports   Giamarchi's assertion 
\cite{Giamarchi_physica} 
 of the irrelevant hopping by umklapp scattering 
 but differs slightly from that by Kishine and Yonemitsu
\cite{Kishine}  who 
 have obtained  the state with reduced but finite interchain hopping.

 Finally  we comment on the  metallic state 
 above the deconfinement transition, 
 which is highly unusual: 
 there is a small Drude weight and a charge gap remaining,  
 while the spin excitations are gapless. 
 The state is similar to that of a doped Hubbard chain
\cite{Mori}. 
 In a simple minded picture 
 single electron transitions between the chains lead 
 to deviations to 1el/unit cell for both chains - 
and thus to a situation also encountered 
 by doping - but whether interchain electron transfer leads 
 to features seen by experiments remains to be seen.  


 This work was partially supported by  a Grant-in-Aid 
 for Scientific  Research  from the Ministry of Education, 
Science, Sports and Culture,(09640429) Japan.

\newpage
\end{multicols}

\end{document}